\documentclass[12pt]{iopart}
\expandafter\let\csname equation*\endcsname\relax
\expandafter\let\csname endequation*\endcsname\relax
\usepackage{adjustbox}
\usepackage{siunitx}
\usepackage{booktabs}
\usepackage{tabularx}
\usepackage{qcircuit}
\usepackage{svg}
\newtheorem{innercustomgeneric}{\customgenericname}
\providecommand{\customgenericname}{}
\newcommand{\newcustomtheorem}[2]{%
  \newenvironment{#1}[1]
  {%
   \renewcommand\customgenericname{#2}%
   \renewcommand\theinnercustomgeneric{##1}%
   \innercustomgeneric
  }
  {\endinnercustomgeneric}
}

\newcustomtheorem{customthm}{Theorem}
\newcustomtheorem{definitionBob}{Definition}

\usepackage{listings}
\usepackage{xcolor}

\definecolor{commentstyle}{HTML}{88846f}
\definecolor{numberstyle}{HTML}{ae81ff}
\definecolor{stringstyle}{HTML}{e6db64}
\definecolor{backcolour}{HTML}{272822}
\definecolor{keywordstyle}{HTML}{f92672}
\lstdefinestyle{mystyle}{
    backgroundcolor=\color{backcolour},  % Background color
    commentstyle=\color{commentstyle}, % comment color
    keywordstyle=\color{keywordstyle}, % e.g. import from for
    stringstyle=\color{stringstyle}, % color of string in python
    identifierstyle=\color{white}, % variable color
    numberstyle=\color{commentstyle}, % <- not responded
    basicstyle=\ttfamily\footnotesize\color{numberstyle}, % bracket, number related
    rulecolor=\color{blue},  
    breakatwhitespace=false,         
    breaklines=true,                 
    captionpos=b,                    
    keepspaces=false,                 
    numbers=left,                    
    numbersep=5pt,                  
    showspaces=false,                
    showstringspaces=false,
    showtabs=false,                  
    tabsize=2
}

\usepackage{color,dsfont} 
\usepackage{graphicx}
\usepackage{subcaption}
\usepackage{ragged2e}
\DeclareCaptionJustification{justified}{\justifying}
\usepackage[normalem]{ulem}
\usepackage{mathrsfs}

\usepackage{mathtools}
\usepackage{float}
\usepackage{verbatim}
\usepackage{latexsym}
\usepackage{setspace}
\usepackage{stmaryrd}
\usepackage{xcolor}
\usepackage{enumitem}
\usepackage{hhline}
\usepackage{xurl}
\usepackage[normalem]{ulem}
\usepackage[colorlinks=true, hyperindex, breaklinks, linkcolor=blue, urlcolor=blue, citecolor=blue]{hyperref} % Physical Review style
\usepackage[capitalise]{cleveref}
\usepackage{multirow}

\def\dya#1{|#1\rangle\!\langle#1|}  % |x><x|
\DeclareUnicodeCharacter{2009}{\,} 

\renewcommand\figurename{FIGURE}
\crefname{equation}{Eq.\!}{Eqs.\!}
\crefname{figure}{Fig.\!}{Figs.\!}
\mathchardef\mhyphen="2D

\RequirePackage{ifthen}

\newboolean{ShowComments}
\setboolean{ShowComments}{true}  % change to true to show remaining colorful author comments. 
\ifthenelse{\boolean{ShowComments}}%
	{
		\newcommand{\ColorComment}[3]{%
				{\colorbox{#1}{\color{white}   \textsf{\textbf{#2}}} \textcolor{#1}{#3}}}%  Colorful box, initials, phrase 

	}%
	{
		\newcommand{\ColorComment}[3]{}%  Do nothing at all

	}%

\definecolor{cocoricolor}{RGB}{200, 110, 180}
\definecolor{michalcolor}{RGB}{255,127,80}
\definecolor{naphanncolor}{RGB}{112, 51, 173}
\definecolor{porametcolor}{RGB}{198,53,39}
\definecolor{rdvcolor}{rgb}{0,0.5,0}
\definecolor{sujincolor}{rgb}{0,0,1}
\definecolor{theerapatcolor}{RGB}{47,123,245}

\definecolor{poompongcolor}{RGB}{0,120,200}

\begin{document}

% \title{Hybrid Strategies for Quantum Communication in Quantum Network of Second Generation Quantum Repeaters}

\title{Hybrid Error-Management Strategies in Quantum Repeater Networks}

\author{Poramet Pathumsoot$^{1,2,*}$, Theerapat Tansuwannont$^{3,4}$, Naphan Benchasattabuse$^{2}$, Ryosuke Satoh$^{2}$, Michal Hajdu\v{s}ek$^{2}$, Poompong Chaiwongkhot$^{1, 5, 6}$, Sujin Suwanna$^{1,*}$, Rodney Van Meter$^{2,*}$}
\medskip
\address{$^{1}$ Optical and Quantum Physics Laboratory, Department of Physics, Faculty of Science, Mahidol University, Bangkok 10400, Thailand\\ $^{2}$ Keio University Shonan Fujisawa Campus, 5322 Endo, Fujisawa, Kanagawa 252-0882, Japan\\ $^{3}$ Institute for Quantum Computing and Department of Physics and Astronomy, University of Waterloo, Waterloo, Ontario, N2L 3G1, Canada\\ $^{4}$ Duke Quantum Center and Department of Electrical and Computer Engineering, Duke University, Durham, NC 27708, USA\\ $^{5}$ National Astronomical Research Institute of Thailand, Chiang Mai, 50180, Thailand\\ $^{6}$ Quantum Technology Foundation (Thailand), Bangkok, Thailand
\\$^*$ Authors to whom any correspondence should be addressed.}
\ead{poramet@sfc.wide.ad.jp, sujin.suw@mahidol.ac.th, \text{and} rdv@sfc.wide.ad.jp}

\begin{abstract}
A quantum network is expected to enhance distributed quantum computing and quantum communication over a long distance while providing unconditional security. As quantum entanglement is essential for a quantum network, major issues from various types of noise and decoherence prevent it from being realized, and research has been intensively active to obtain optimal configurations for a quantum network. In this work, we address the performance of a quantum network capable of quantum error correction and entanglement purification. Our results show that one should distribute Bell pairs as fast as possible while balancing the deployment of fidelity enhancement. We also show suitable hybrid strategies in quantum cryptography tasks under some noise regimes that need to use purification and quantum error correction together. Our results suggest that using purification to distribute high fidelity Bell pairs and preserving them for application using quantum error correction is a promising way to achieve a near-term quantum network for secure communication.
\end{abstract}

%\pacs{03.67.Pp}

\maketitle

%Main contents.

% \input{Chap_1_intoduction}

% \input{Chap_2_background}

% \input{Chap_3_results}

% \input{Chap_4_simulation_results}

% \input{Chap_5_discussion}

\section{Introduction} \label{section:introduction}

%\PP{Key message: Our results suggest that using purification to distribute high-fidelity Bell pairs and preserve them for application using quantum error correction is promising way to achieve near-term quantum network for secure communication.}

%\PP{This section will describe a motivation of quantum network starting from quantum computer point of view to quantum cryptography. As current technologies still faulty, careful design of entanglement needed for quantum network is important to save to cost and achieve useful result. Our simulation show that memory time of qubit is an important factor that contribute to a large portion of final fidelity, thus we design an entanglement distribution such that the high quality entanglement will be delivered using entanglement purification while using quantum error correction to maintain its quality. The summary of each subsection will be briefly described.}

\begin{figure}
    \centering
    \includegraphics[width=0.95\textwidth]{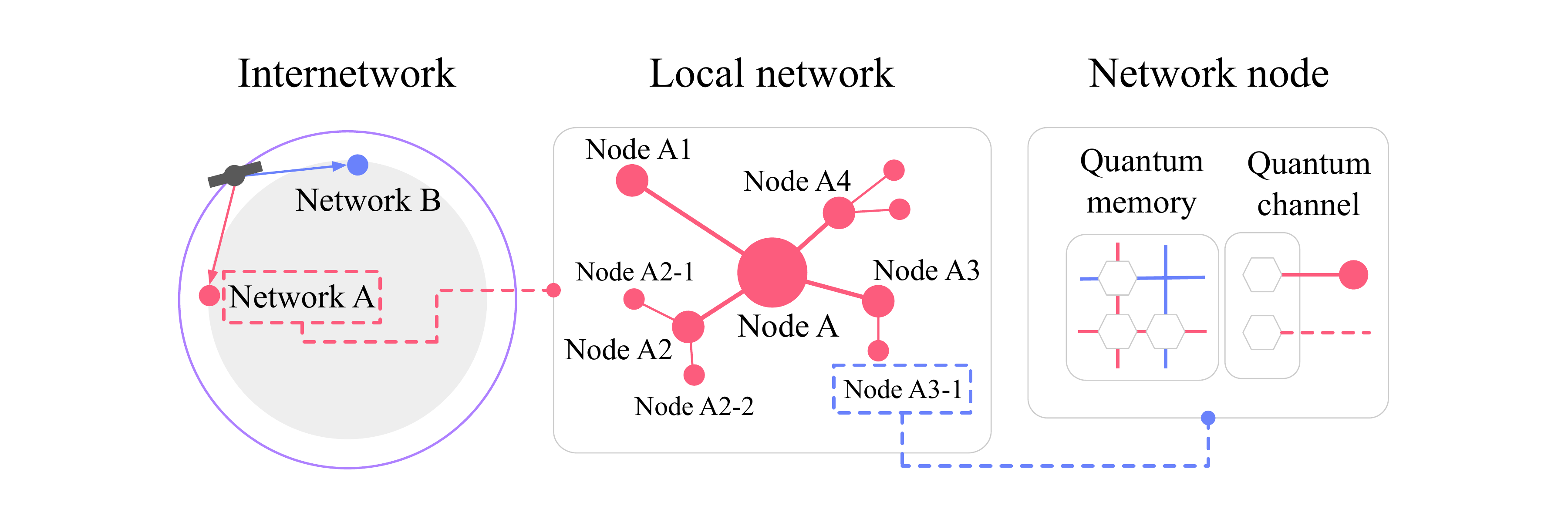}
    \caption{Concept of a worldwide quantum Internet. Inter-continental quantum communication may be realized with the aid of a network of satellites \cite{Khatri2021, Mol2023}, which act as long-distance Bell pair distributors for main hubs such as Node A and Node B. These hubs slice the long-distance Bell pairs with short-range entangled states generated by the local networks. }
    \label{fig:QNCopcept}
\end{figure}

Quantum communication is at the forefront of the second quantum revolution~\cite{dowling2003quantum}.
It is expected to be instrumental in unlocking the true potential of quantum computers~\cite{arute2019quantum,zhong2020quantum,madsen2022quantum} in the form of distributed quantum computing~\cite{cuomo2020towards,caleffi2022distributed}.
Other equally exciting applications include secure quantum key distribution~\cite{ekert1991quantum,yin2017satellite,sasaki2017quantum}, improved arrays of sensing devices~\cite{gottesman2012longer,bartlett2007reference,ilo-okeke2018remote}, and secure private cloud services~\cite{fitzsimons2017private,broadbent2009universal,hayashi2018self}.
These applications will rely on quantum networks~\cite{vanmeter2014quantum,I-D.irtf-qirg-principles-11} to distribute entanglement between their participants in an efficient, fair, and reliable fashion.
Following the evolution path of the classical Internet, it is expected that quantum networks themselves will one day be connected and lead to the ultimate goal of quantum communication in the form of a global quantum internet~\cite{kimble2008quantum,wehner2018quantum,satoh2021attacking} as illustrated in Fig.~\ref{fig:QNCopcept}.

Similar to classical networks, quantum networks also utilize light as the primary information carrier between nodes of the network.
Unlike their classical counterparts, arbitrary quantum states cannot be copied faithfully~\cite{park1970concept,wootters1982single,dieks1982communication}, making classical signal amplification schemes impossible to apply \cite{chia2019phase}.
Coupled with the fact that quantum nodes exchange information via single photons, attenuation becomes a major obstacle to scaling of quantum networks.
One approach to overcoming the exponential attenuation in fiber is to segment the entire connection between the quantum nodes via the use of \emph{quantum repeaters}~\cite{briegel1998quantum,azuma2022quantum}.
Link-level entanglement is distributed between neighboring repeater stations, which then splice them via \emph{entanglement swapping}~\cite{PhysRevLett.71.4287, pan1998experimental} into an end-to-end entangled connection.
% Due to the probabilistic nature of entanglement swapping with linear optics, the end-to-end entanglement distribution rate decreases polynomially with the distance.
This repeat-until-success approach means that the end-to-end entanglement distribution rate decreases polynomially.

% The main obstruction needed to overcome is a noise which attack quantum information as its sent along the distant path. Unlike a classical network, where signals can be modified and amplified with intermediate stations along the way, a quantum state cannot be copied and re-sent. Hence, distributing quantum resources such as entanglement and qubits over a long distance network presents an enormous challenge for the signal boosters must also be quantum systems. This requires quantum memories and light-matter interface. An analogue of a classical amplifier or signal booster is called a \emph{quantum repeater} proposed as an intermediate node to reduce the distance quantum information has to travel. Nevertheless, noises are still present in the system despite shortening distance. The capability of a quantum repeater is crucial in the quest to achieve a better performance of a network. 

%\textcolor{red}{literature review}

Besides entanglement swapping, another important role of the quantum repeaters is to participate in error management.
Decoherence in quantum memories, imperfect quantum gate operations, and measurement errors all introduce unwanted \emph{operational errors}, which must be mitigated in order to ensure end-to-end entanglement remains above the desired threshold fidelity required by the application.

Error management schemes can be classified into three generations~\cite{muralidharan2016optimal}.
The first generation (1G) of repeaters relies on entanglement purification schemes~\cite{briegel1998quantum,bennett1996purification,pan2001entanglement,dur2007entanglement} to detect errors on physical qubits.
The need for two-way classical communication between distant nodes in the network makes this generation of repeaters unsuitable for long-distance quantum communication.
However, due to their relatively modest hardware requirements and low resource overhead, they are expected to be the primary method of error management in early implementations of small quantum networks.
The second generation (2G) of repeaters~\cite{jiang2009quantum,fowler2010surface} uses quantum error correction (QEC) to both detect and correct errors.
This generation of repeaters avoids the issue of two-way communication and is therefore more suitable for long-distance quantum communication, albeit at the price of stricter hardware requirements and larger resource overhead.
The third generation (3G) of quantum repeaters~\cite{muralidharan2014ultrafast,munro2012quantum} relies on quantum error correction to correct both loss and operational errors and does not require entanglement swapping or pre-shared long-distance entanglement as a means of communication.
Thus, 3G repeaters place more strict requirements on operational fidelity and place more complex demands on the hardware than 2G repeaters.
% Further extension of a quantum repeater is proposed by \cite{Muralidharan2016}, separating quantum repeater into three generations, discriminated by their approaches of handling noises. A first generation uses the entanglement purification; a second generation uses quantum error correction both required entanglement swapping to establish long-range communication, while a third generation enables sending quantum information directly through a quantum channel as a one-way communication.

The performance of individual repeater generations has been analyzed and contrasted in previous reports; see for examples Refs.~\cite{muralidharan2016optimal,dur1999quantum,fujii2009entanglement,jansen2022enumerating}.
Apart from the generation of the repeater network, the network performance also depends on the entanglement distribution policy \cite{khatri2021policies}.
In this work, we present a comprehensive study of a number of entanglement distribution strategies and compare their performance under realistic noise conditions in terms of end-to-end fidelity and throughput.

We show by simulation that using either purification or QEC alone is not enough to produce end-to-end Bell pairs with sufficiently high fidelity above a certain threshold in the noise regime we investigated.
Instead of deploying QEC and purification techniques separately, we utilize them together by introducing a hybrid strategy, which we refer to as \emph{purified encoding} (PE).
PE first increases the fidelity of physical Bell pairs by using purification and then encodes a logical Bell pair.
We investigate two variations of this hybrid strategy. First, we perform PE immediately after producing a physical Bell pair at the link level. Second, we employ PE at the end nodes. %Each technique could be used before the other.
We identify the error parameter regime,  quantum gate errors, and measurement errors, where PE outperforms strategies based purely on 1G or 2G.
% As a cost of quantum network infrastructure might be astronomical high, a simulator is a necessary tool for assessing the performance of the network to effectively design and compare strategies for conducting real experiments. Our simulator can keep track of Pauli errors of the system instead of those of the quantum states themselves. With a direct fidelity estimate using the stabilizer counting method, this allows us to simulate the network efficiently without spending exponentially cost of memory space as traditional method. Even with the disadvantages of restricting to only Pauli errors, we could evaluate performance of quantum error correcting code with ease. 

Our paper is organized as follows.
In \cref{section:fromTheoryToSimulation}, we give a brief overview of the relevant concepts concerning quantum networks, quantum repeaters, entanglement swapping, entanglement purification, and quantum error correction.
We then provide a detailed description of our simulator, its assumptions, noise models, method for fidelity evaluation, as well as validation with theoretical models in simple cases where analytical results are obtained. 
We discuss, in \cref{section:setting}, the simulation assumptions and the parameter values. 
In \cref{section:result}, we present the simulation results comparing the performance of various strategies for distributing Bell pairs in various regimes of noises.
In this work, we consider noise from depolarizing errors in a quantum channel, memory time of a qubit, photon loss, gate errors, and measurement errors.
In a linear chain of nodes, the distance between the two end nodes is fixed while we vary the number of quantum repeaters in between.
Conclusions and discussion of our results are summarized in \cref{section:discussion}.

% \Figure[t!](topskip=0pt, botskip=0pt, midskip=0pt){contents/quantum-internet.pdf}
% {Concept of world-wide quantum internet. Inter-continental quantum communication may realize by an aid of network of satellites, which act as a Bell pair distributor for main hubs such as Node A and Node B in figure. The hubs provide an entanglement to the nodes in their network using their shared Bell pairs.\label{fig:QNCopcept}}

\section{From Theory to Simulation} \label{section:fromTheoryToSimulation}

%\PP{Discuss about method to distribute Bell pair in the literature. Quantum repeater is categorized into three generation as in [Muralidharan2016]. This should softly introduce noises in quantum network. Each generation will be briefly discuss, 0G will introduce entanglement swapping. Then 1G will introduce entanglement purification along with the protocol that we will used in the work, Ss-Dp. 2G will introduce with Quantum error correction, stressing on Steane code as we use this code in the work. From the needed to simulate large among of qubit, we discuss our method via error-basis model, show noise channel and evaluation against analytical expression.}

In this section, we briefly introduce the essential elements used in a quantum network. 
We then discuss the error models we used and verify the output of our simulation by comparing it with analytically derived expressions for simple scenarios.

\subsection{Quantum Network: Theoretical Framework}\label{subsec:theoryPOV}

While it is possible to distribute Bell pairs directly via optical fiber, it is not efficient since the probability of success decreases exponentially as the distance increases.
This exponential attenuation can be overcome by the use of a quantum repeater~\cite{Briegel1998}.
This way, Bell pairs need only be distributed over short distances connecting individual repeaters.
The main function of a quantum repeater is then to splice this link-level entanglement into entangled states shared between far-away nodes in the network through entanglement swapping, as well as managing the inevitable operational errors via purification or quantum error correction. We will now give a brief overview of these basic functions of a quantum repeater.

% A quantum repeater, proposed in~\cite{Briegel1998}, is an intermediate quantum node placed between two destination nodes.
% Each end node first shares matter-matter Bell pairs with a quantum repeater; then the Bell measurement is performed on the qubits at the repeater to produce long-range Bell pair between the two end nodes. Unfortunately, photon loss and decoherence are not all the noise that exist in the system; a quantum network also suffers from quantum gate errors, measurement errors, and depolarizing noise in a quantum channel. To tackle the problems, a network of quantum repeaters is extended and categorized into first generation (1G), second generation (2G) and third generation (3G) with different configurations to minimize errors and maintain high fidelity, as proposed in \cite{muralidharan2016optimal}. Before proceeding to state the precise description of each generation of quantum repeaters, we will first briefly discuss an entanglement swapping protocol as it is used in both 1G and 2G network of quantum repeaters.

\textit{Entanglement Swapping} is a form of quantum teleportation where the entanglement between a pair of qubits is transferred to a new qubit.
It can be demonstrated on a simple example of two Bell pairs, one shared between qubits A and B, and the other between qubits C and D, respectively.
Measurement of qubits B and C in the Bell basis will project previously independent qubits A and D onto one of the maximally entangled state.
The randomness of the measurement outcome can be compensated by applying a correction operator conditioned on the two classical bits representing the measurement results.

\begin{figure}
        \centering
        \includegraphics[width=0.9\columnwidth]{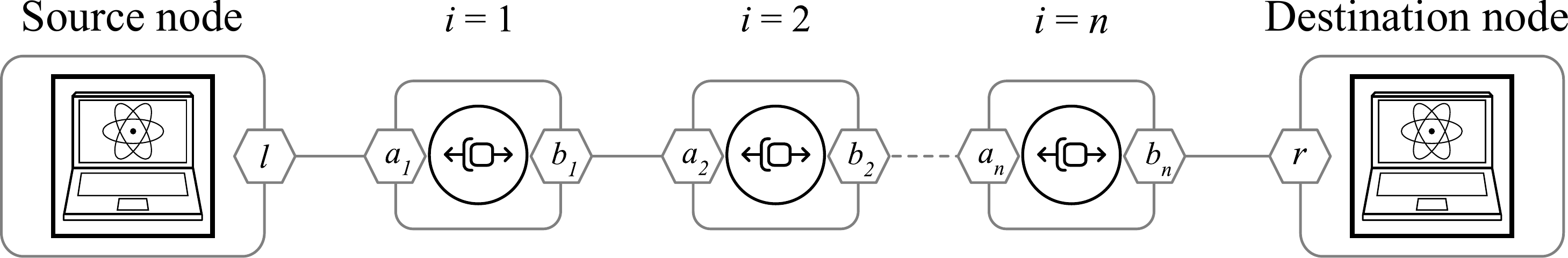}
        \caption{Linear chain of quantum repeaters from the source (initiator) node to the destination (responder) node. The figure uses icons proposed in Ref.~\cite{9951258} with permission.}
        \label{fig:LinearChain}
    \end{figure}

Entanglement swapping is at the heart of the entanglement-splicing function of a quantum repeater as pictured in~\cref{fig:LinearChain}.
Two end nodes, here referred to as the Source and Destination nodes, can share an entangled Bell pair with the aid of a chain of $n$ repeaters.
The qubit that is held by the Source node's qubit is labeled as $l$, while the Destination node's qubit is labeled as $r$.
Repeater $i$ is in possession of at least two qubits, labeled as $a_i$ and $b_i$, respectively.
Each repeater measures its two qubits in the Bell basis, producing two classical bits, which we also label as $a_i$ and $b_i$.
The total correction needed to be applied by either the Source node or the Destination node is then $\Pi_i X^{b_i}Z^{a_i}$, which can be simplified by further use of the anticommutation property of the Pauli matrices $XZ=-ZX$. However, our implementation of entanglement swapping is not simultaneously executed, but in a round-based approach; please refer to \ref{subsec:simulation}.

% In the simplest case where $n = 1$, two Bell pairs and three quantum nodes will involve in the protocol. To have the initiator node (A) share a Bell pair with the responder node (C), the intermediate node (B) performs BSM on two qubits from each Bell pair that it holds. Measurement results $a_{1} \text{and } b_{1}$ are then sent to the initiator or responder nodes, which inform the end node of which operation is needed to perform on the qubit in order to correct the state to the Bell state.

In reality, distributed Bell pairs will be affected by various types of noise.
Assuming that the probability of any of the Pauli matrices affecting the entangled state is equal, the distributed state of fidelity $F$ can be written as
\begin{equation}
    \label{eq:densityBell}
    \rho = F \dya{\Phi^{+}} + \frac{1 - F}{3} (\dya{\Phi^{-}} + \dya{\Psi^{+}} +\dya{\Psi^{-}}),
\end{equation}
where $|\Phi^{\pm}\rangle = (|00\rangle \pm |11\rangle) / \sqrt{2}$,  $|\Psi^{\pm}\rangle = ( |01\rangle \pm |10\rangle) / \sqrt{2}$.
Performing entanglement swapping on two such mixed states, shared between qubit pairs A-B and C-D with fidelities $F_{AB}$ and $F_{CD}$, respectively, will result in a mixed state shared between the qubits A-D with fidelity \cite{doi:https://doi.org/10.1002/9781118648919.ch10}
\begin{equation}
    \label{eq:fidelityES}
    F_{AD} = F_{AB}F_{CD} + \frac{(1 - F_{AB})(1 - F_{CD})}{3}.
\end{equation}

\begin{figure}[t]
    \centering
    \includegraphics[width=0.7\textwidth]{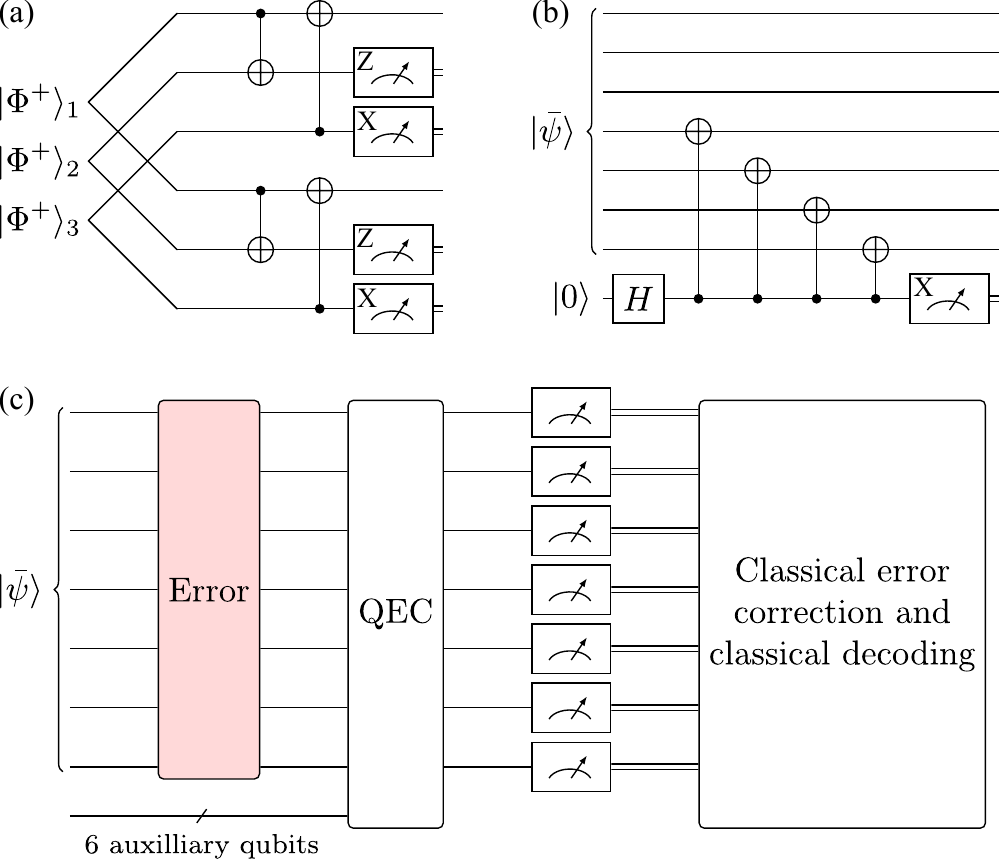}
    \caption{(a) 1G repeaters use entanglement purification as a form of error detection. We make use of the Ss-Dp protocol that requires two ancillary Bell pairs to check whether the original Bell pair is affected by $X$ or $Z$ Pauli errors. (b) Example of a syndrome measurement of the stabilizer generator $g_3=IIIXXXX$. (c) The full procedure to measure a logical qubit. The measurement basis of the logical qubit depends on the measurement basis of the physical qubits.}
    \label{fig:error_management}
\end{figure}

\textit{Entanglement Purification:} We now turn our attention to how quantum repeaters manage errors.
We begin with 1G quantum repeaters which use entanglement purification \cite{dur1999quantum, Deutsch1996} to detect errors affecting the entangled states.
If no error is detected, our confidence in the state's quality increases leading to improved fidelity.
We use the single selection, double purification (Ss-Dp) protocol \cite{fujii2009entanglement,matsuo2019quantum, Matsuo2019} where three Bell pairs are required for this protocol, as depicted in Fig.~\ref{fig:error_management}(a).
One auxiliary Bell pair is used to measure information about the Pauli $X$ error, and another one for measuring the Pauli $Z$ error.

\textit{Quantum Error Correction:} 2G and 3G quantum repeaters rely on quantum error correction to both detect and correct errors.
We use the $[[7, 1, 3]]$ Steane code \cite{Steane1996}, which uses seven physical qubits to encode one logical qubit.
%, and can independently correct the Pauli $X$ and Pauli $Z$ errors using 3 additional qubits each.
The Steane code is a stabilizer code \cite{Gottesman97} which has a set of stabilizer generators given as follows: $g_1 = XIXIXIX$, $g_2 = IXXIIXX$, $g_3 = IIIXXXX$, $g_4 = ZIZIZIZ$, $g_5 = IZZIIZZ$, $g_6 = IIIZZZZ$ and logical operators $\Bar{X} = XXXXXXX$, and $\Bar{Z} = ZZZZZZZ$.
Measuring these generators results in an error syndrome which identifies an error on the physical qubits.
We denote the physical qubits of a logical qubit $A$ by lower case $a_i$ where $i=1,...,7$. One important feature of the Steane code is that the logical Clifford gates can be implemented transversally; logical Hadamard and logical $S$ gates can be implemented by applying corresponding physical gates on all physical qubits, while a logical CNOT gate, $\text{CNOT}(A,B)$, can be implemented between two blocks of code by applying physical $\text{CNOT}(a_i,b_i)$ between all seven pairs of physical qubits.

%As the Steane code belongs to the CSS code family \cite{CSS}, the syndrome measurements of $Z$ and $X$ errors can be performed simultaneously.\TT{How? Isn't this true for any stabilizer code?}

As the Steane code belongs to the CSS code family \cite{Steane1996, CSS}, one can correct an $X$-type (or a $Z$-type) error using only the measurement results obtained from $Z$-type (or $X$-type) generators. In this work, the error syndrome is obtained by measuring all generators concurrently using three auxiliary qubits per type, hence six auxiliary qubits in total. Figure~\ref{fig:error_management}(b) shows an example of stabilizer generator $g_3$.
Simultaneous generator measurements can be beneficial to a matter qubit which requires long measurement and re-initialization time, as one does not need to reset the qubit between each generator measurement. However, the process comes with the cost of additional auxiliary qubits. Once the error syndrome is obtained, quantum error correction can be done by applying a Pauli operator corresponding to the syndrome to the physical qubits. The quantum error correction is performed before and after each logical quantum gate.

The measurement readout of a logical qubit using the Steane code can be obtained in a number of ways~\cite{devitt2013quantum}.
In this work, a logical qubit measurement in an $X, Y,$ or $Z$ basis is done by measuring all physical qubits in the corresponding Pauli basis.
If there is no error on the physical qubits and the measurements are perfect, the bitstring of measurement outcomes will be one of the codewords of the $[7,4,3]$ classical Hamming code, thus the logical state can be found by applying classical decoding to the bitstring.
On the other hand, if there is an error on one physical qubit or a single-qubit measurement error (but not both), the bitstring of measurement outcomes will differ from a codeword of the $[7,4,3]$ classical Hamming code by one bit.
In this case, the logical state can be obtained by applying classical error correction to the bitstring followed by classical decoding.
Here we also perform quantum error correction before each logical qubit measurement. The whole procedure is illustrated in \cref{fig:error_management}(c).
However, it should be pointed out that this procedure is not fault tolerant. This means that any errors, including memory errors, errors arising from faulty gates, and measurement errors, can propagate to other qubits depending on the subsequent operations.

\subsection{Error Models and Assumptions} \label{sec:ErrorModelAssumption}

We now provide details of the error models used in our work. Our design and assumptions of the error propagation follow those in Ref.\cite{Satoh2021}, which tracks information about whether a qubit has been affected by a Pauli $Z$ and/or $X$ error along with a time stamp when it was initiated.
We now summarize the error sources which contribute towards the degradation of the fidelity of the end-to-end entangled state.

We begin by considering the link-level entanglement distribution involving the entanglement swapping of two light-matter Bell pairs that results in a matter-matter Bell pair.
The \emph{depolarizing channel} is applied to each matter qubit, where the depolarizing action on a quantum state $\rho$ is defined as
    \begin{equation}\label{eq:depolarizingChannel}
        % Depolarizing error
        \mathcal{E}_{\text{depo}}(\rho) = (1 - p_{\text{depo}}) \rho + \frac{ p_{\text{depo}}}{3} (X\rho X^{\dagger} + Y\rho Y^{\dagger} + Z\rho Z^{\dagger}),
    \end{equation} 
where $p_{\text{depo}}$ denotes a depolarizing error probability.

% Gate error
\emph{Gate error} is implemented in the form of a depolarizing channel acting on the qubit with the probability controlled by the parameter $\lambda_{\text{gate}}$. For a single-qubit gate $G_1$, the quantum channel with gate errors acting on $\rho$ was implemented as

%Likewise, \emph{gate error} is also implemented in the form of depolarizing channel acting on a qubit with probability $p_{\text{gate}}$, and $P_i$ are Pauli matrices. For a single-qubit gate $G_1$, the gate error on $\rho$ is defined as
    \begin{equation} \label{eq:oneQubitChannel}
        % One qubit gate error channel
        \mathcal{E}_{1}(\rho) = (1 - \lambda_{\text{gate}})G_1 \rho G_1^{\dagger} + \frac{\lambda_{\text{gate}} }{4}\sum_{i\in\{I, X, Y, Z\}} P_{i} G_1\rho G_1^{\dagger}P_{i}^{\dagger},
    \end{equation}
where $P_i$ represents Pauli operators. Similarly, the quantum channel for a two-qubit gate $G_2$ with gate errors acting on $\rho$ is defined as
\begin{equation} \label{eq:twoQubitChannel}
        % Two qubit gate error channel
        \mathcal{E}_{2}(\rho) = (1 - \lambda_{\text{gate}})G_2 \rho G_2^{\dagger} + \frac{\lambda_{\text{gate}} }{16}\sum_{i, j\in\{I, X, Y, Z\}} (P_{i}\otimes P_{j}) G_2\rho G_2^{\dagger}(P_{i}^{\dagger}\otimes P_{j}^{\dagger}).
\end{equation}    
We use a single parameter $\lambda_{\text{gate}}$ to control the error rates for both single- and two-qubit gates.
Our implementation takes into account the fact that two-qubit gates usually have higher error rates.
A given value of $\lambda_{\text{gate}}$ results in the single-qubit gate error rate of $p_1 = \frac{3 \lambda_{\text{gate}}}{4}$ and the two-qubit gate error rate of $p_2 = \frac{15 \lambda_{\text{gate}}}{16}$. Thus, the quantum channels corresponding to one-qubit and two-qubit gates, respectively, are
\begin{align} 
    % \label{eq:oneQubitChannel}
    % One qubit gate error channel
    \mathcal{E}_1(\rho) & = (1 - \frac{4}{3}p_1)G_1 \rho G_1^{\dagger} + \frac{p_1 }{3}\sum_{i\in\{I, X, Y, Z\}} P_{i} G_1\rho G_1^{\dagger}P_{i}^{\dagger}, \\
    %\label{eq:twoQubitChannel}
    % Two qubit gate error channel
    \mathcal{E}_2(\rho) & = (1 - \frac{16}{15}p_2)G_2 \rho G_2^{\dagger} + \frac{p_2 }{15}\sum_{i, j\in\{I, X, Y, Z\}} (P_{i}\otimes P_{j}) G_2\rho G_2^{\dagger}(P_{i}^{\dagger}\otimes P_{j}^{\dagger}).
\end{align} 
Ideally, the two error rates should be independent, but for the sake of keeping the simulation time reasonable while still reflecting the current state of quantum hardware, we have opted for this approach.

% Memory error
%We do not implement memory error as a relaxation process.
Similarly to Ref.\cite{Hartmann2007}, our simulator assumes \emph{memory error} to be a depolarizing error with probability $p_{\text{mem}}$ defined as
    \begin{equation} \label{eq:memoryModel}
        p_{\text{mem}} = 1- \left(\frac{3e^{-t/\tau} + 1}{4} \right) = \frac{3}{4}(1-e^{-t/\tau}),
    \end{equation} 
where $t$ denotes the time of measurement, and $\tau$ the memory lifetime. The quantum channel with memory errors acting on $\rho$ is 
\begin{equation}\label{eq:memoryChannel}
        \mathcal{E}_{\text{mem}}(\rho) = (1-p_{\text{mem}}) \rho + \frac{ p_{\text{mem}}}{3} \left(X\rho X^{\dagger} + Y\rho Y^{\dagger} + Z\rho Z^{\dagger} \right),
\end{equation}
and it is applied to the qubit immediately before any measurement.

(Although \cref{eq:depolarizingChannel} and \cref{eq:memoryChannel} are similar, $\mathcal{E}_{\text{depo}}(\rho)$ is time-independent and applies only when link-level Bell pair is generated, while $\mathcal{E}_{\text{mem}}(\rho)$ is time-dependent is applies only before any measurement on a physical qubit.)

% Measurement error
The measurement process is essential in a quantum network, especially with Bell state measurement (BSM) and quantum teleportation, and it can incur errors on a measured qubit. We model \emph{measurement error} as a bit flip on the classical readout.
% As our simulator yields only bit-flip errors in a measure basis, the measurement error could be easily implemented.
The simulator flips the measurement outcome with probability $p_{\text{meas}}$.
More details about the error model and how it is applied can be found in \ref{subsec:simulation}.

\subsection{Fidelity Evaluation} \label{subsec:FidelityEvaluation}
The main goal of our simulator is to evaluate the quality of the distributed end-to-end Bell pairs.
State tomography could be used to estimate the final state but would incur a heavy computational tax.
Our primary metric of interest is the fidelity of the distributed state with respect to a certain Bell state, which can be estimated without the full knowledge of the density matrix in certain cases~\cite{flammia2011direct}.

The fidelity of a Bell pair can be estimated by measuring its stabilizer operators. We will now demonstrate how this method works by expressing the Bell pair $|\Phi^+\rangle$ in terms of the Pauli operators,
\begin{equation}\label{eq:Bellpair}
    |\Phi^{+}\rangle \langle \Phi^{+}| = \frac{1}{4} \left( II + XX - YY + ZZ \right).
\end{equation}
An arbitrary two-qubit state $\rho$ can be written as
\begin{equation}
    \rho = \frac{1}{4} \sum_{i,j\in I,X,Y,Z} \langle P_i \otimes P_j \rangle P_i \otimes P_j,
\end{equation}
where $\langle P_i \otimes P_j\rangle = \text{Tr} \left[ (P_i \otimes P_j) \rho \right]$.
Using the properties of the Pauli matrices, it can be readily shown that the fidelity of this arbitrary state with respect to the Bell pair is
\begin{equation} \label{eq:fidelityBell}
    F(\rho, |\Phi^{+}\rangle \langle \Phi^{+}|) = \langle \Phi^{+}| \rho |\Phi^{+}\rangle
    = \frac{1}{4}\left(1 + \langle XX \rangle - \langle YY \rangle + \langle ZZ \rangle \right).
\end{equation}

Our simulator tracks the probability that a qubit is affected by a particular Pauli error.
The correlations $\langle P_i \otimes P_i \rangle$ in \cref{eq:fidelityBell} can be obtained directly from these error probabilities in the following way.
We check whether $P_i \otimes P_i$ commutes or anticommutes with a particular Pauli error operator.
If it commutes, then we leave its probability as is. If it anticommutes, we multiply its error probability by $-1$.
Finally, we sum all the probabilities and renormalize them by the number of qubits, which in this case is two.
This can be done during the quantum state evolution without having to collect data and constructing a quantum state from a tomographic process.
We refer to this method of computing the fidelity as \emph{direct fidelity estimate}; see \ref{subsec:simulation}.
    
\subsection{Validation with Theoretical Model} \label{sec:Validation}
In this subsection, we verify the direct estimation of the fidelity used in our simulation by comparing it with analytically derived expressions for certain simple scenarios.
The setting we consider is that of two end nodes trying to share a Bell pair with the help of a single quantum repeater placed half-way between the two end nodes.
The case of unequal link lengths has been investigated in Ref.\cite{pathumsoot2021optimizing}.
This is the same scenario discussed in \cref{subsec:theoryPOV}, but this time due to the assumption of equidistant separation of the end nodes and the repeater, the link-level entangled states will have the same fidelity, $F_{AB} = F_{CD} = F_{\text{link}}$.
% Considering the depolarizing channel defined in \cref{eq:depolarizingChannel}, it is possible to derive analytical expressions for the fidelity of a distributed Bell pair after entanglement swapping as well as after purification.

\begin{figure}
        \centering
        \includegraphics[width=\columnwidth]{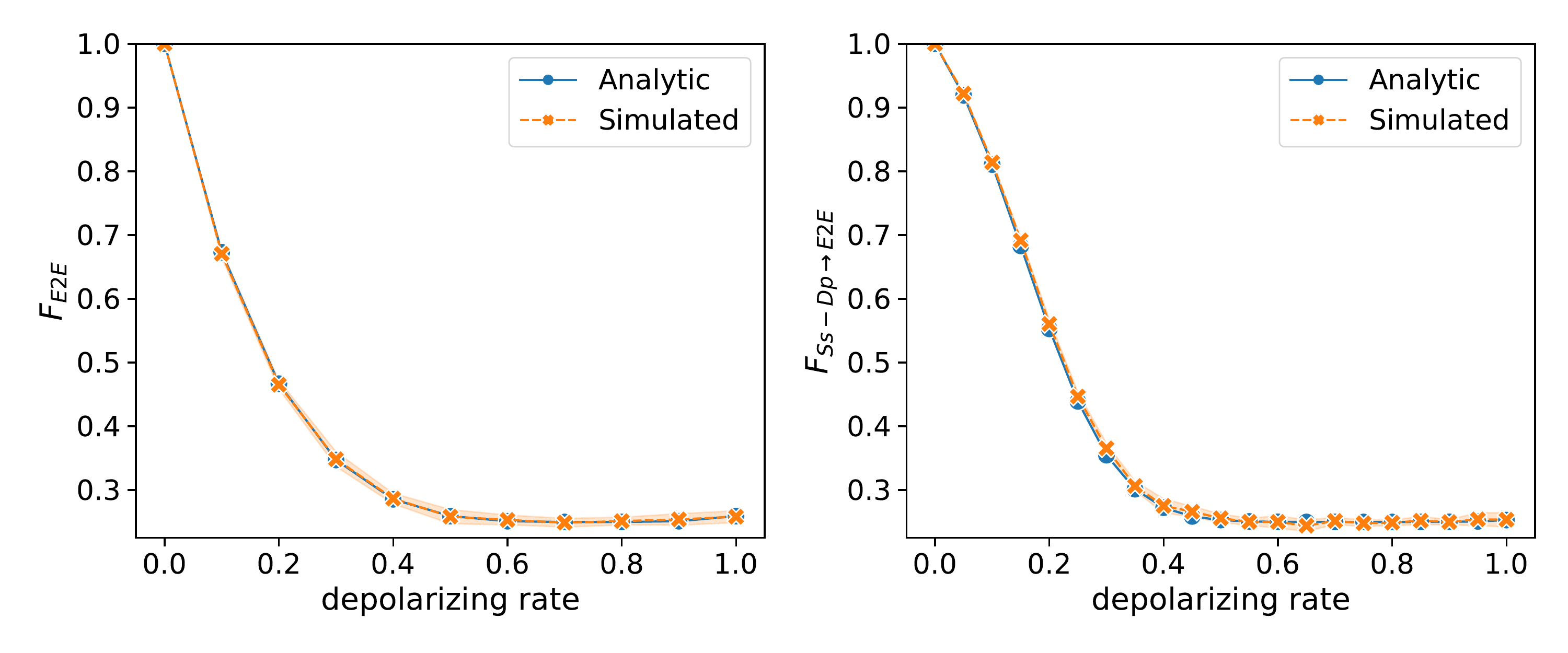}
        \caption{(Left) Fidelity of the end-to-end Bell pair after entanglement swapping using noisy Bell pairs passing through a depolarizing channel. Good agreement of analytical and simulated results  is evident. Each point is an averaged value of 10 trajectories, and \num{9000} measurements for each trajectory. (Right) Fidelity of the end-to-end Bell pair after entanglement swapping using purified noisy Bell pairs (by the Ss-Dp protocol) passing through depolarizing channel. Likewise, each point is a an average value of 10 trajectories, and \num{9000} measurements for each trajectory.}
        \label{fig:0G_1G_model}
    \end{figure}

%\naphann{Waiting for Poramet's confirmation. Appendix B seems to suggest this is correct. MM-link is used but the equation suggests that both ends initialize qubits at the same time thus undergoing a waiting time of $2L/c$. Also, we probably should add a note that our fiber only has loss error and no Pauli error.} \PP{I added explanation in the appendix B about that, but for Puali error, does depolarizing channel count?}
Considering the depolarizing channel in \cref{eq:depolarizingChannel}, the link-level fidelity with respect to the state $|\Phi^+\rangle$ is
    \begin{equation}\label{eq:withoutpurificaiton}
        F_{\text{link}} = \frac{ p_{\text{depo}}^2}{3} + (1 -  p_{\text{depo}})^2.
    \end{equation}
% From \cref{eq:densityBell}, we identify that $F = F_{\text{raw}}$.
Using \cref{eq:fidelityES}, we obtain the fidelity $ F_{\text{E2E}}$ of the end-to-end Bell pair after entanglement swapping:
\begin{equation}
    F_{\text{E2E}} = \left(\frac{p_{\text{depo}}^{2}}{3} + \left(1 - p_{\text{depo}}\right)^{2}\right)^{2} + \frac{1}{3} \left(1 - \frac{p_{\text{depo}}^{2}}{3} - \left(1 - p_{\text{depo}}\right)^{2} \right)^{2}.
\end{equation}
We benchmark the results generated by our simulator with the analytic expression of this model, as shown in the left panel of \cref{fig:0G_1G_model} with varying values of $p_{\text{depo}}$.
    
    % \begin{figure}
    %     \centering
    %     \includegraphics[width=\columnwidth]{0G_model.png}
    %     \caption{Fidelity of analytic and simulated Bell pair after entanglement swapping using noisy Bell pairs passing through a depolarizing channel. Each point is a mean value of 10 trajectories, and \num{9000} measurements for each trajectory.}
    %     \label{fig:0G_model}
    % \end{figure}

Moreover, the analytical fidelity of the Bell pair resulting from the Ss-Dp protocol corresponding to \cref{eq:densityBell} is given by
    \begin{equation} \label{eq:SsDpFidelity}
        F_{\text{Ss-Dp}} = \frac{14 F_{\text{link}}^{2} - 7 F_{\text{link}} + 2}{16 F_{\text{link}}^{2} - 14 F_{\text{link}} + 7},
    \end{equation}
where $F_{\text{link}}$ is the fidelity of the link-level physical Bell pair given by \cref{eq:depolarizingChannel}; see \ref{sec:derivation}. The right panel of Fig.~\ref{fig:0G_1G_model} shows the end-to-end fidelity of obtained by first applying the Ss-Dp purification protocol to both links before executing entanglement swapping.
We denote this fidelity by $F_{\text{Ss-Dp} \rightarrow \text{E2E}}$.
The analytic expression for this fidelity is lengthy, and it can be obtained by using \cref{eq:withoutpurificaiton} and \cref{eq:SsDpFidelity} along with \cref{eq:fidelityES}.
The simulated and analytical results show excellent agreement.
Small discrepancies occur in both cases because our direct fidelity estimate is a statistical measure, paid in exchange for lower computational cost.
    
    % \begin{figure}
    %     \centering
    %     \includegraphics[width=\columnwidth]{1G_model.png}
    %     \caption{Fidelity of analytic and simulated Bell pair after entanglement swapping using purified noisy Bell pairs passing through depolarizing channel. Each point is a mean value of 10 trajectories, and \num{9000} measurements for each trajectory.}
    %     \label{fig:1G_model}
    % \end{figure}

\section{Simulation Setting and Parameters} \label{section:setting}

In this section, we motivate the configuration of our simulator before discussing the entanglement distribution strategies analyzed in the rest of this article.
The strategy to establish long-range entanglement depends strongly on the noise regime as well as the demands of the intended application.
These requirements are often formulated in terms of threshold fidelity \cite{Perseguers2010}.
For example, a QKD application that is robust against the faked Bell state attack requires the end-to-end fidelity of at least 0.83~\cite{Sajeed2019BrightlightDC}.
We use this threshold fidelity as a convenient and useful goal for the considered strategies and show that the purely 1G or purely 2G distribution strategies are unable to reach it under our simulation conditions.
On the other hand, our new hybrid strategies are capable of surpassing this threshold fidelity for some noise regimes which we identify.

% A quantum communication strategy to establish long-range entanglement in a quantum network depends on the noise regime. A secure network, a quantum state fidelity must meet a required threshold \cite{Perseguers2010}, which could be a different value for each different task.
% In this section, we first state the configuration of our simulation and its motivation, where we assume the fidelity threshold of a QKD system to be 0.83, which is minimal requirement against faked Bell state attack \cite{Sajeed2019BrightlightDC}. Secondly, we introduce some strategies considered in this work, and discuss the results on their performances.

We now give details of all the fixed parameters that we use in our simulations.
\begin{enumerate}
    \item We consider a linear chain of repeaters with the distance between the two end nodes being fixed at \num{100} \si{\km}. The link lengths are all equal but the number of repeaters can be varied.
    \item The speed of light in fiber is assumed to be constant at \num{300000} \si{\km/\s}.
    %it serves as a delay of light travelling and also to check the correctness of timing of the simulation during development. 
    \item The depolarizing probability is fixed at $p_{\text{depo}}=0.025$. It should be noted that the entanglement generation over a distance of 50 km using $^{40}$Ca$^{+}$ ion in Ref.\cite{Krutyanskiy2019} reported a Bell pair of fidelity of 0.86 (simulation) and 0.86 $\pm$ 0.03 (experiment), where the simulation accounted for the effect of measured background counts only. Compared to our work, including only depolarizing channel applied directly on the ideal Bell pair, setting $p_{\text{depo}}$ to be approximately 0.0736 results in the output fidelity of approximately 0.86.
    \item The memory lifetime of $\tau = 10 \text{ ms}$. This generous lifetime is typical in a number of physical systems, particularly in ion traps~\cite{Wang2021} and NV centers in diamond~\cite{pompili2021realization}. 
    \item We assumed the effective loss rate to be \num{0.30} \si{dB/km}.
    This is a conservative value; optical fibers with substantially lower attenuation have been used in experiments.
    This value is sufficient to affect the quality of qubits that suffer from a long waiting time.

%\HR{We did not assume photon loss rates in separate media such as fibre, atmosphere, or free-space.} The combined effects of memory time and photon loss are \HR{sufficient} to affect the quality of qubits that suffer from a long waiting time. 
%\PCc{saying what was excluded from the model might cause more trouble than help. Suggest edit: We assumed effect loss rate of our model to be 0.3 dB/km. This has taken into account...list here..}  x
\end{enumerate} 
The above parameters have typical, or worse than typical, values compared to the experimental ones reported in Refs.~\cite{Hofmann2005, Liao2017, Chen1989, Valentini2021AnalysisOP, Abasifard2023}, and are kept fixed in the strategy evaluation. Variable parameters for optimization are the gate error parameter $\lambda_{\text{gate}} \in \{0.0000,0.0005,0.0010,0.0015,0.0020\}$, the measurement error $p_{\text{meas}} \in \{0.0000,0.0025,0.0050,0.0075,0.0100\}$, and the number of hops $h_s \in \{2, 4, 8\}$.

\begin{figure*}[t]
    \centering
    \includegraphics[width=\textwidth]{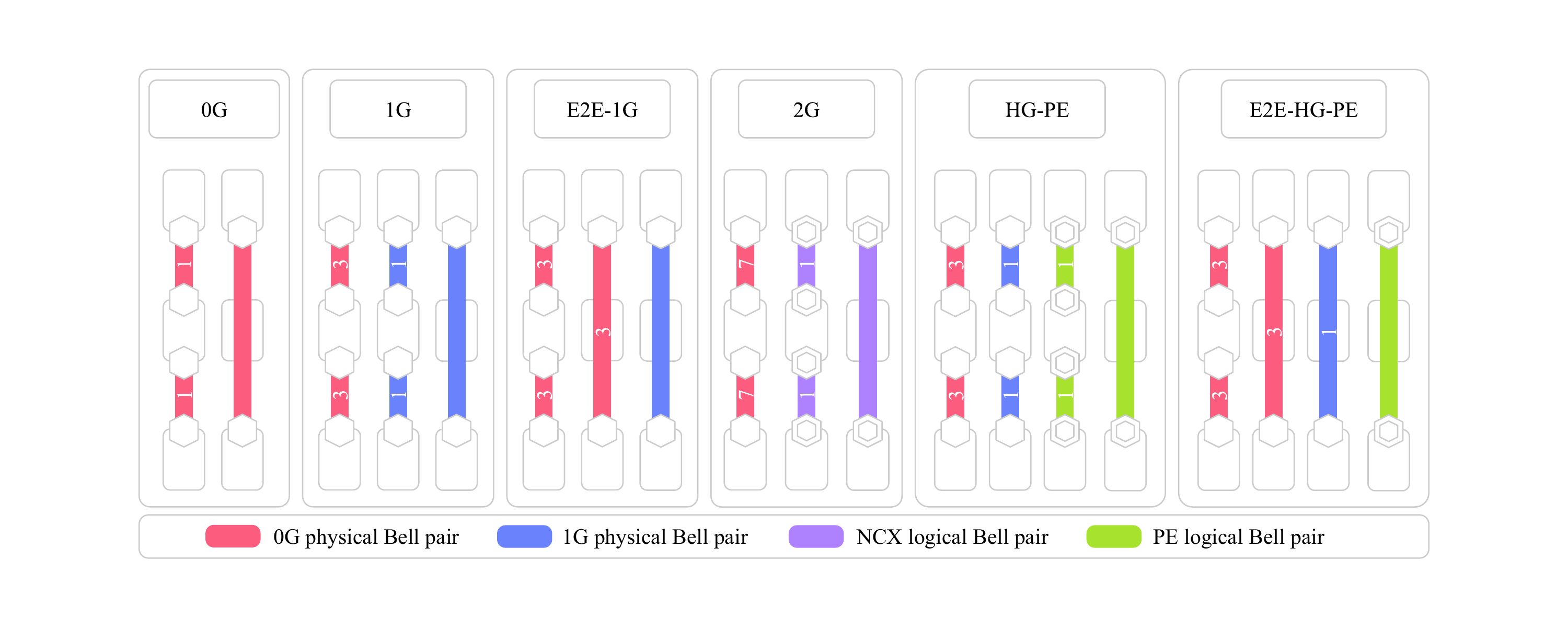}
    \caption{Entanglement distribution strategies. Each strategy shows Bell pairs drawn vertically, with end nodes at the top and bottom and repeaters (if any) in between. Actions are shown left to right, beginning with raw link-level Bell pairs and ending with the high-fidelity end-to-end Bell pair. The numbers on the colored links indicate the number of Bell pairs from the previous round consumed in the current round.
    %\textbf{0G}: Immediately after Bell pairs on both sides of the quantum repeater are ready, entanglement swapping is performed until a Bell pair is distributed to the end nodes. \textbf{1G}: Ss-Dp protocol is employed on link-level Bell pairs first, then entanglement swapping is performed to until the Bell pair is distributed to the end nodes. \textbf{E2E-1G}: Similar to \textbf{0G}, but at the end nodes, Ss-Dp protocol is performed to produced the final Bell pair. \textbf{2G}: At the link-level, 7 Bell pairs will be used to perform Non-local CNOT gate between logical qubits on adjacency nodes, then entanglement swapping is performed on the logical Bell pairs until the end nodes receive a logical Bell pair.  \textbf{HG-PE}: At the link-level, Ss-Dp is performed on physical Bell pairs, then it will be encoded to a logical Bell pair directly without using a Non-local CNOT gate, entanglement swapping is then perform until end nodes produced a logical Bell pair. \textbf{E2E-HG-PE}: Similarly to \textbf{0G} and \textbf{HG-PE}, physical Bell pairs are distributed to the end nodes using \textbf{0G}, then those Bell pairs will be purified and used to encoded logical Bell pair as in \textbf{HG-PE} to produce the final logical Bell pair.
    }
    \label{fig:strategies}
\end{figure*}

We now discuss six different entanglement distribution strategies used in our analysis, as illustrated in \cref{fig:strategies}. The most basic strategy relies on entanglement swapping only, without utilizing any purification or quantum error correction. We denote this strategy by [\textbf{0G}]. It serves as a reference for the performance of other strategies. 
%\HR{We remark that our 0G strategy is equivalent to 2G (NC) in Ref.\cite{muralidharan2016optimal}. \PP{I'm not sure about this statement.}\PCc{agreed. This statement might cause misunderstanding to our claims and does not directly contributed to the main result. Suggest toning down or remove.}} 

We now consider strategies for 1G repeater networks using the Ss-Dp purification protocol.
There are two variations of the 1G-only strategy.
First, the simple 1G strategy, denoted by [\textbf{1G}], is to perform purification at the link-level only.
Entanglement swapping is performed using the successfully purified Bell pairs to produce the final Bell pair between the end nodes.
The second variation, which we call [\textbf{E2E-1G}] strategy, is to first perform entanglement swapping to generate the Bell pairs for the end nodes.
Then the Ss-Dp protocol is performed to produce the final purified Bell pair (hence, the name E2E which stands for end-to-end).
Since producing a purified link-level Bell pair is expected to take longer for [\textbf{1G}] hence delaying the swap, the [\textbf{E2E-1G}] alternative is considered to determine whether it can increase the final end-to-end fidelity and throughput compared to [\textbf{1G}]. 

For the pure [\textbf{2G}] strategy, we use the standard seven-qubit Steane code to encode quantum information into a logical state.
The strategy is to first distributed to link-level physical qubits and used to perform logical non-local CNOT (NCX) gates on logical qubits in the state $|\bar{+}\rangle$ in the left node and $|\bar{0}\rangle$ in the right node to generate one link-level logical Bell pair. After that, the entanglement swapping is used to distribute the logical Bell pairs to the end nodes. QEC is applied before every measurement on a logical qubit, and before and after the application of every logical gate. 

Similarly to the investigation in Ref.~\cite{PhysRevA.93.042338}, it is useful to consider using purification and QEC altogether.
We first enhance the 1G strategies with QEC, which makes the strategies a hybrid between 1G and 2G.
After the Ss-Dp protocol, the purified Bell pairs are used for encoding logical qubits, effectively a logical Bell pair. These hybrid strategies are denoted by [\textbf{HG-PE}], and [\textbf{E2E-HG-PE}], which are the extensions of [\textbf{1G}] and [\textbf{E2E-1G}], respectively. A summary of different strategies used in this work is depicted in \cref{fig:strategies}.

\section{Results} \label{section:result}

\begin{figure*}[t]
    \centering
    \includegraphics[width=0.95\textwidth]{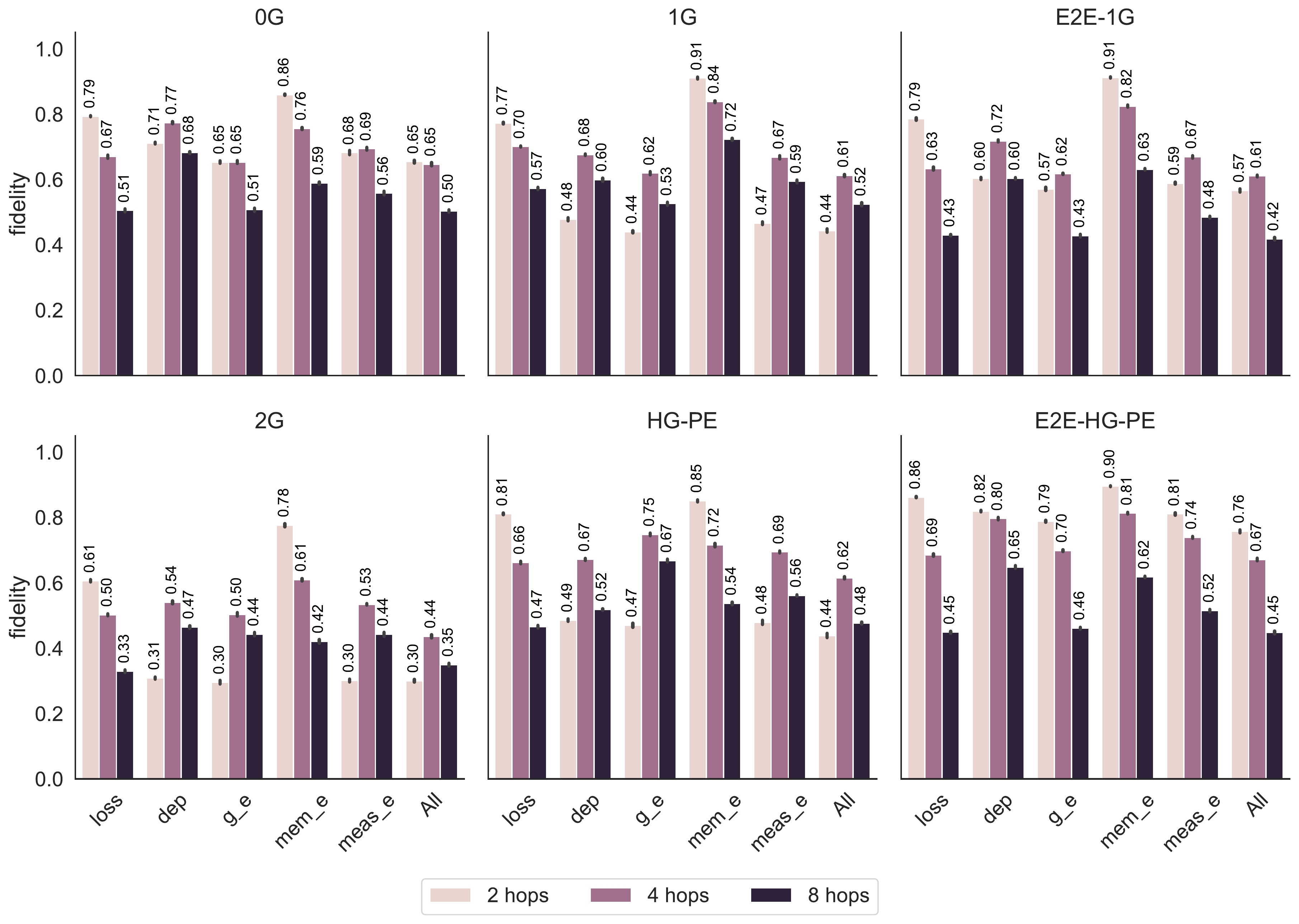}
    \caption{End-to-end fidelity in the presence of noise. The horizontal axis indicates that all noise sources are present except the labelled one, where loss is photon loss, \texttt{dep} is depolarizing channel, \texttt{g\_e} is the gate error channel, \texttt{mem\_e} is the memory error channel, and \texttt{meas\_e} is the measurement error channel. While \texttt{All} means that all channels were included in the simulation. Each case is subject to depolarizing error of 2.5\%, gate error of 0.1\%, measurement error of 1\%, memory error of 0.01\si{s}, and loss of 0.3\si{dB/km}.}
    \label{fig:NoiseTesting}
\end{figure*}

Let us now present the simulation results.
In Fig.~\ref{fig:NoiseTesting}, we investigate how the end-to-end fidelity is affected by both the number of hops as well as the role individual sources of noise play in its degradation.
We ran separate simulations for every instance of turning off a particular source of noise, as indicated by the labels on the horizontal axis.
Holding the probability of depolarizing error $p_{\text{depo}}$ constant at 2.5\%, the fidelity of Bell pairs generated between adjacent nodes is independent of the distance.
However, for some strategies, the fidelity does not always decrease as the number of hops increases.
This seemingly counter-intuitive observation is a direct consequence of the fact that the total distance between end nodes is fixed.
Increasing the number of hops, therefore shortens the link length.
Combined with finite memory lifetime, this results in the observed behavior, where simply increasing the number of hops can lead to higher end-to-end fidelity.

\begin{figure*}[t]
    \centering
    \includegraphics[width=\textwidth]{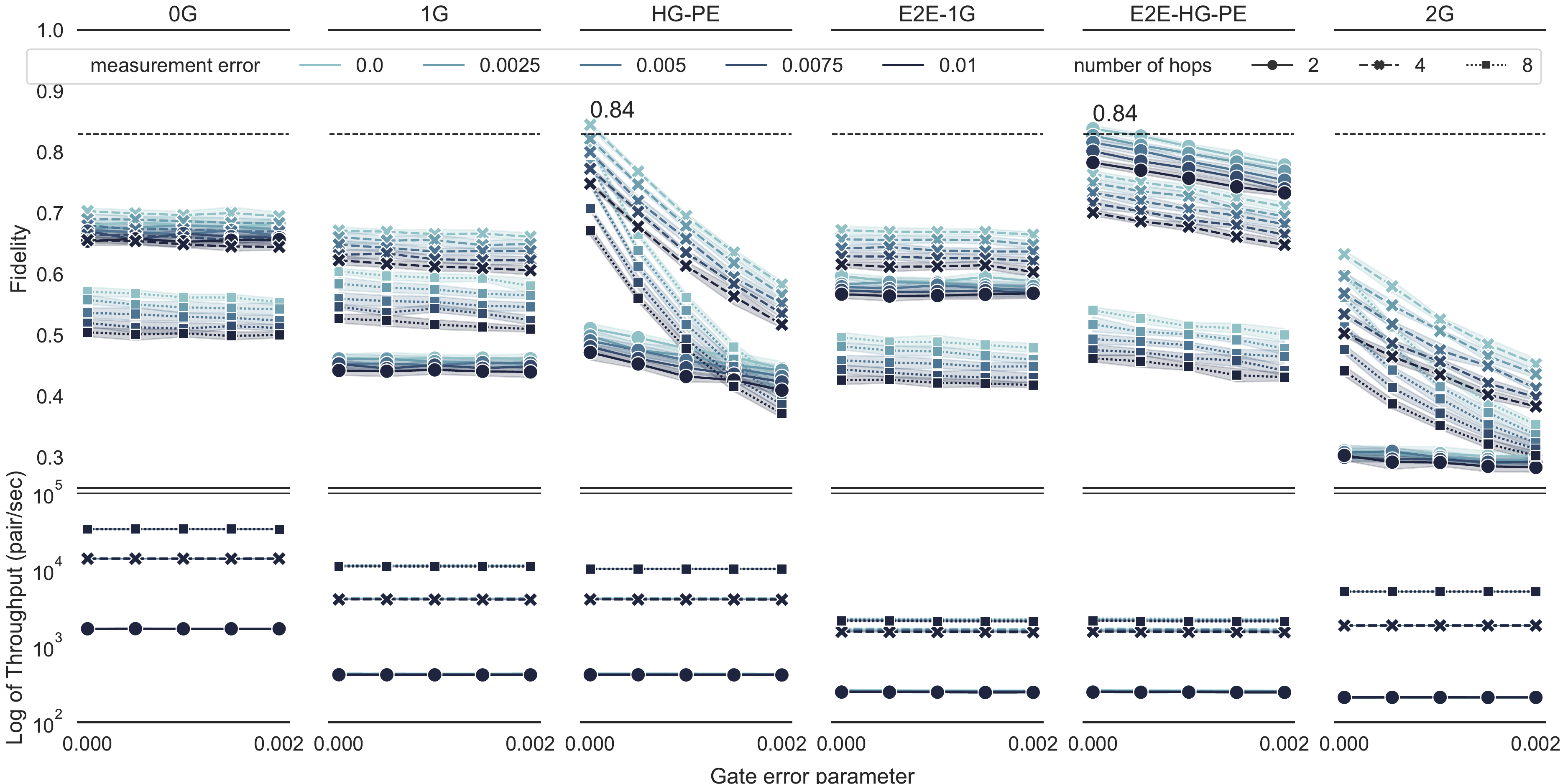}
    
    \caption{The upper plots show the fidelity of the end-to-end Bell pair yielded from each strategy with varying gate error parameter $\lambda_{\text{gate}} \in \{0.0000,0.0005,0.0010,0.0015,0.0020\}$, measurement error $m_e \in \{0.0000,0.0025,0.0050,0.0075,0.0100\}$ and the number of hops $h_s \in \{2, 4, 8\}$. The dashed line is the reference fidelity of value 0.83. The lower plots show the throughput of each strategy on a logarithmic scale.}
    \label{fig:fidelity}
\end{figure*}

Setting the memory lifetime to be infinite, the fidelity decreases as the number of hops increases.
From \cref{fig:fidelity}, simply distributing the Bell pairs fast is not sufficient as 0G also fails to reach the threshold fidelity of 0.83.
Hence, the need for noise suppression techniques is warranted.
However, it is evident that 1G also could not produce Bell pairs that reach this fidelity threshold.
Even its extension E2E-1G suffers from short memory time such that the purification could not offer any advantage over 0G at all.
Even worse, 2G, which requires more physical resources, effectively prolongs the waiting time and yields Bell pairs of fidelity smaller than other strategies.
From the aforementioned results, we regard short memory lifetime as a main problem to be tackled, consistent with recent report \cite{Mol2023}. 

As pointed out in Ref.~\cite{PhysRevA.93.042338}, it is useful to consider using purification and QEC together.
A key reason that we consider a hybridization between 1G and 2G is that the logical qubit measurement method described in \cref{subsec:theoryPOV} (which is only applicable to CSS codes such as the Steane code) allows additional classical error correction after measurement. 
Thus, in addition to measurement error, hybridization can also correct errors due to short memory lifetime as well.
As shown in \cref{fig:fidelity}, the [\textbf{HG-PE}] and [\textbf{E2E-HG-PE}] strategies offer relatively higher fidelity than the other strategies.
Specifically, 2-hops of [\textbf{E2E-HG-PE}] outperforms the other strategies in the high error rate regime.
It also reaches the desired threshold in the absence of gate errors.
The drawback of [\textbf{HG-PE}] is that it suffers from gate errors as it uses many more gate operations, similar to other 2G strategies.
Thus, the fidelity falls quickly when gate errors increase.
The [\textbf{E2E-HG-PE}] strategy, on the other hand, uses QEC in the later stages of the protocol.
Hence, it does not induce unnecessary gate errors.
In terms of resource efficiency, the [\textbf{E2E-HG-PE}] strategy also consumes fewer Bell pairs compared to [\textbf{HG-PE}] and does not demand the intermediate repeaters to be 1G or 2G.
%With a strategy that uses [\textbf{0G}] to deliver end-to-end Bell pairs and uses QEC to help with error correction \TT{sounds repetitive}, 
%Our results suggest that distributing high-quality Bell pairs as a starting point and then preserving them over a long period of time with QEC yields high fidelity Bell pair for application usage.
In essence, to deliver end-to-end Bell pairs with QEC, our results suggest that distributing high quality Bell pairs, then preserving them over a long period of time with QEC yields high fidelity Bell pairs for application usage.

\section{Discussion} \label{section:discussion}

Our results show that the failure to realize high fidelity quantum communication is a result from short memory lifetime. 
It is generally not a good strategy to simply increase the number of hops along the path to increase the end-to-end Bell pair production rate as it unavoidably induces more errors in the system due to imperfections of the applied operations.
A lower number of quantum repeaters also leads to the decoherence of qubits due to increased waiting times during which quantum memories are locked.

Our proposed solution is to use a hybrid approach.
\cref{fig:fidelity} shows that in order to deliver high fidelity Bell pairs to the application, it is desirable to distribute physical Bell pairs with 0G between end nodes followed by purification to increase their fidelity, before finally encoding them as logical Bell pairs.
% As the Bell pairs can be generated beforehand and consumed by the application on demand, the key idea of delivering long live high fidelity, as \cref{fig:fidelity} suggests, is to distribute physical Bell pairs directly with 0G to the end nodes, then use purification to increase the fidelity of a physical Bell pair to be encoded as a logical Bell pair.
End-to-end Bell pairs can be generated by the quantum network in advance and then consumed on demand by the application.
Using purification and QEC helps preserve the fidelity of the end-to-end Bell pair until it is needed by the application.
Compared to [\textbf{E2E-1G}], using a pure strategy in a 1G or 2G network is not sufficient to suppress errors in the system.

It would be interesting to consider [\textbf{0G}] to distribute Bell pairs for encoding at the destination nodes, or other purification schemes \cite{fujii2009entanglement, dur2007epa}.
As our memory error model is quite simple, more advanced models might reveal other interesting behavior of the network.

\appendix

\section{Fidelity derivation}\label{sec:derivation}

Here, we use algorithms to generate the analytical expressions of \cref{eq:withoutpurificaiton} and \cref{eq:SsDpFidelity} coded in \texttt{python} using \texttt{sympy} and \texttt{pandas} for visualization. The algorithms combinatorically calculates the probability of all possible combinations of depolarizing errors that could occur on qubits to be used in operations. With probabilities of all combination calculated, additional operations are performed such as entanglement purification. Finally, the probabilities of scenarios that will result in an expected Bell pairs are summed together and normalized by the summation of the probabilities of all events of interest. The result is effectively the fidelity of a Bell pair. Please refer to our \url{https://github.com/PorametPat/qwanta/blob/main/experiments/exp_id4_QubitModule/exp_id4_Purification.ipynb} for codes and results. 

\section{Details of simulation}\label{subsec:simulation}

To simulate the complex behavior of the quantum network, time delay and resource management, we use \texttt{SimPy}, the discrete-event simulation package written in Python. For noise simulation, as large quantum systems needed to be simulated, similarly to Stabilizer formalism, we use error-basis model \cite{Satoh2021} which collects only the information about the noise of each qubit thus allow us to efficiently simulate the system. Although this is not the representation of a quantum state, the choice is preferred as it is relatively easier to implement and the information of the noise is enough to be used for the fidelity discussed in \cref{subsec:FidelityEvaluation}.

In the following, we explain the implementation of operation in quantum network in our discrete-event simulator. Entanglement whether physical or logical Bell pair is referred to as a resource. We categorize operations in quantum networks into the elementary process as described in detail below.   %\PCc{what is the purpose of the word 'now'? Would it be beter using more generic tone like "In our work we expand the implementation of....."}

% Explain each process
 % 1. Generate physical resource
\emph{Generate physical resource} Generation of physical resource or link-level entanglement generation is a process that takes two external physical qubits from adjacent nodes and creates a Bell pair. Consider Node 1 and Node 2, after the \emph{Emitter} sub-process receives a fresh physical qubit from each node. We assume that the qubit $q_1$ at Node 1 will emit a photon qubit to the Bell state analyzer (BSA) at Node 2. $q_1$ is initialized (setting time reference to the beginning of the simulation), and the time required for light to travel from Node 1 to Node 2, $t_{\text{delay}}$, is calculated and used to delay the next step. The BSA is incorporated into one of the nodes, so that photons are traveling the link distance in only one direction. This architecture is referred to as "memory-memory" (MM) link~\cite{jones2016design}. In the following sub-process, \emph{Qubit} $q_2$ is initialized, and it emits a photon to its BSA. The probability of detecting both photons, $p_{\text{success}} = 10^{-\xi\cdot d/10}$, is used to determine the link-generation result, where $\xi$ is loss measured in unit of \si{\dB/\km}, and $d$ is the relative distance in \si{km} between Node 1 and Node 2. The result is then sent from the Node sub-process to  Node 1, which will delay the next step by $t_{\text{delay}}$ in a form of classical message. If the attempt is success, the depolarizing channel as describe in \cref{eq:depolarizingChannel} will applied to both matter qubits, and those two qubits will be registered as a physical Bell pair for other subsequent processes. On the other hand, if the attempt fails, both qubits are reset, and put back for subsequent attempts. 

% 2. Generate logical resource
A process for generating logical Bell pairs is \emph{Generate logical resource}. There are two approaches to generating logical Bell pairs. One method is to use non-local CNOT gates to perform logical CNOT gates on both logical qubits at each node. To perform a non-local CNOT gate in this method, we need seven physical Bell pairs. After acquiring seven physical Bell pairs and seven qubits to encode logical qubits on each node, the protocol begins, where the logical qubits are initialized at each node. At Node 1, the CNOT gates are performed using each physical qubit as a control and one end of the physical Bell pair as a target; this end of the physical Bell pairs are measured in the $Z$ basis, and then measurement results are send to Node 2 with time delay $t_{\text{delay}}$. With measurement results, error is propagate and correction operator is perform with probability of $1/2$. At Node 2, CNOT gates are performed with a qubit from a Bell pair as a control qubit and physical qubit used for encoding logical qubit as a target qubit, on control qubit, Hadamard gates are performed then, measuring in Z-basis. Measurement results are sent, inducing a delay of $t_{\text{delay}}$, with measurement results, errors are propagate and correction operators are perform with probability of $1/2$ each. The qubits composing Bell pairs used for this process are then reset and returned to be used later. The logical Bell pair is then registered as a new logical resource shared between Node 1 and Node 2. Another method uses one physical Bell pair as an input for logical qubit encoding, and the two logical qubits will result in a logical Bell pair. After getting one physical Bell pair shared between Node 1 and Node 2, and the six physical qubits for encoding from each node, the encoding process starts. Each node will use one-end of the Bell pair and six physical qubits to encode the logical qubit. The resulting logical Bell pair will be registered as a new logical resource. 
% 3. Purification
\emph{Entanglement Purification} is a process that purifies one physical Bell pair referred to as a resource pair with auxiliary Bell pairs. We use the Ss-Dp purification protocol \cite{matsuo2019quantum, Matsuo2019, fujii2009entanglement} which requires two auxiliary Bell pairs. After getting three physical Bell pairs at each node, we apply CNOT using a resource qubit as a control and the first auxiliary qubit as a target, then we apply CNOT using the second auxiliary qubit as a control and a resource qubit as a target. We then measure the first auxiliary qubit in the $Z$ basis and the second auxiliary qubit in the $X$ basis, and the measurement result will be shared to determine the next step. If measurement result from both qubits of the pair agree, then the resource pair is accepted as a purified resource, and is registered in the resource pool. Also, the auxiliary qubits are reset and returned to the fundamental pool. Otherwise, all qubits are reset and put back in the fundamental pool. 
        
% 4. Entanglement swapping
%\naphann{It seems to me that the correction operations are only applied at the end nodes as mentioned in~\ref{subsec:theoryPOV} , so it is not clear to me about the waiting time differences between each strategy. I think it should be mentioned a little clearer here.} \PP{Added in the bottom.}
\emph{Entanglement swapping} is a process that creates a Bell pair from two Bell pairs. In this situation, we assume that Node 1 shares a Bell pair with Swapper, and Swapper shares a Bell pair with Node 2. After getting two Bell pairs, one from each node, we perform the following step, depending on whether the logical Bell pair or physical Bell pair is received. For a physical Bell pair, Swapper proceeds to change the measurement basis for each qubit from each pair to a Bell basis with CNOT and Hadamard gate. Swapper then measures both qubits in the $Z$ basis, and returns the measured qubits to the fundamental resource. For a logical qubit, each node will request six qubits for syndrome measurement, and we perform quantum error correction (QEC) on both logical qubits, once before and one after the application of CNOT gate. QEC is performed again on the logical qubit after the Hadamard gate is applied, but we do not apply QEC on another logical qubit. Both qubits are then measured in the $Z$ basis. After acquiring the measurement results, the delay induced is the time for sending a classical message to Node 1 and Node 2. However, only the longer time delay will be used in a subsequent process. As errors can propagate from measurement results, correction operators are performed assuming that four possible combination measurement results are equally likely. The correction and error propagation is done only to qubits in Node 2. After the correction, a new logical Bell pair will then be registered in the resource pool.
%\PP{I added the following to clarify the process further.} 
Regarding the swapping strategy used, we implement a round-based approach. For 2 hops, we need only one entanglement swapping, the strategy is trivial. For 4 hops, in the first stage, we perform entanglement swapping on nodes (0, 1, 2) and nodes (2, 3, 4) simultaneously, and then in the second stage, Bell pairs from the first stage are used to perform entanglement swapping on node (0, 2, 4). Lastly, for 8 hops, in the first stage, we perform protocol on (0, 1, 2), (2, 3 ,4), (4, 5, 6), (6, 7, 8), Bell pairs produced are used for second stage on nodes (0, 2, 4) and (4, 6, 8), then final stage with nodes (0, 4, 8). We also note that this entanglement swapping is implemented for all memory qubits only. For entanglement swapping of memory-photon Bell pairs is already done via measurement using BSA. 
        
% 5. Fidelity estimation
\emph{Fidelity estimation} is a process which uses many Bell pairs to estimate the fidelity of a resulting qubit with a Bell state. The process uses one Bell pair at a time, where the measurement results will be stored and used to calculate the fidelity after number of measurements required, $n_\text{meas}$, is reached. After acquiring a Bell pair, if it is a logical Bell pair, we perform the following step. Each node requests six auxiliary qubits and uses QEC. After syndrome measurement, the auxiliary qubits will be reset and returned to the fundamental pool. We measure each qubit in each basis in the following order, $XX, YY, \text{and } ZZ$ for $\lfloor n_\text{meas}/3 \rfloor$ each. After measurement, qubits are reset and put back in the fundamental pool. If the measurement occurs $n_\text{meas}$ times, then the fidelity is calculated as described in section \cref{subsec:FidelityEvaluation}.
We used the \texttt{dill} package \cite{McKerns2010Pathos:Computing, McKerns2012BuildingScience} for saving and loading data in our simulation.

\section*{Acknowledgement}
This research has received funding support from the NSRF via the Program Management Unit for Human Resources \& Institutional Development, Research and Innovation [grant number B05F650024]. This material is based upon work supported by the Air Force Office of Scientific Research under award number FA2386-19-1-4038.

\section*{References}
\bibliographystyle{unsrt}
\bibliography{references}

%\newpage

%\appendix

%\input{App_Jul14}

\end{document}